\documentclass [aps,amsmath,amssymb,superscriptaddress,preprint]{revtex4}
\usepackage{graphicx}

\def\gsim{\mathrel{\rlap{\lower4pt\hbox{\hskip1pt$\sim$}}
    \raise1pt\hbox{$>$}}}       

\begin{document}

\title{Discrete Hilbert Space, the Born Rule, and Quantum Gravity}  

\author{Stephen~D.~H.~Hsu} \email{hsu@msu.edu}
\affiliation{Department of Physics and Astronomy\\ Michigan State University \\ \bigskip \bigskip}

\begin{abstract}
\bigskip
Quantum gravitational effects suggest a minimal length, or spacetime interval, of order the Planck length. This in turn suggests that Hilbert space itself may be discrete rather than continuous. One implication is that quantum states with norm below some very small threshold do not exist. The exclusion of what Everett referred to as maverick branches is necessary for the emergence of the Born Rule in no collapse quantum mechanics. We discuss this in the context of quantum gravity, showing that discrete models (such as simplicial or lattice quantum gravity) indeed suggest a discrete Hilbert space with minimum norm. These considerations are related to the ultimate level of fine-graining found in decoherent histories (of spacetime geometry plus matter fields) produced by quantum gravity.
\end{abstract}


\maketitle

\date{today}


\section{Introduction: Minimal Length and Discrete Hilbert Space}

Quantization of gravity may lead to the existence of a minimal length, or (via relativistic covariance) a minimal spacetime interval \cite{Calmet:2004mp}. Quantum gravity is expected to produce strong fluctuations of the metric at the Planck scale, precluding any semi-classical notion of shorter distances. In fact, no macroscopic experiment can be sensitive to discreteness of position on scales less than the Planck length. Any device (such as
an interferometer, e.g., as used by LIGO) capable of such resolution would be so massive that it would have already collapsed into a black hole. 

That the classical metric $g_{\mu \nu}$ might dissolve into quantum fluctuations at the Planck scale is by now a familiar idea. Less well appreciated is that this might have consequences for the nature of Hilbert space itself \cite{BHZ1}. We can argue as follows. Consider an experiment which takes place in a spacetime region of extent $L$. Given the short distance cutoff at the Planck length, $l_p$, the number of degrees of freedom relevant to the experiment is itself finite. Imagine that the experiment measures the state of a single qubit -- the orientation of a spin:
\begin{equation}
\vert \psi \rangle = \cos \theta \, \vert + \rangle ~+~ e^{i \phi} \sin \theta \, \vert - \rangle~~~.
\end{equation}
For a given $L$, the number of distinct configurations of the experimental apparatus (i.e., the number of distinct quantum operators represented by the possible measurements) is bounded above. Thus the number of distinct spin orientations (qubit states which are eigenstates of the measurement operator) that can be resolved is also bounded above. Physics can therefore be described by a discretized Hilbert space in which the angles $( \theta, \phi )$ are discrete and take on only a finite (but presumably very large) number of values. Holography (another aspect of quantum gravity) provides a stronger bound on the scale of discreteness: the total entropy of the measurement apparatus is bounded above (which limits its configuration and accuracy of read out) by the boundary area rather than the volume of the region in Planck units.

Under the assumptions described above, no experiment can exclude the possibility that Hilbert space is discrete and finite dimensional, i.e.,  
\begin{equation}
\label{psisum}
\vert \psi \rangle = \sum c_n \, \vert n \rangle ~~~
\end{equation}
where I. the values of the coefficients $c_n$ are only defined to some finite accuracy -- they are {\it not} continuous complex parameters, and II. the sum itself is finite.

A physicist who has simulated quantum phenomena on a classical computer may not find properties I and II very shocking. What we describe above as fundamental consequences of quantum gravity are approximations made out of necessity in everyday computation. 

Two states are indistinguishable if $\vert \psi - \psi' \vert < \epsilon$, for some very small $\epsilon$. Assuming the holographic bound above, then for a single qubit $\epsilon \sim L^{-1}$ (note we will use Planck units in what follows) \cite{BHZ2}. If $L$ is the size of the visible universe, then $\epsilon < 10^{-61}$.

\bigskip

Equivalently, given two (normalized) quantum states $\psi$ and $\phi$, and $\lambda$ {\it sufficiently small},
\begin{equation}
\label{snap}
\vert \psi \rangle + \lambda \, \vert \phi \rangle \approx \vert \psi \rangle ~~~,   
\end{equation}
where $\, \approx \,$ means {\it physically equivalent to} in some fundamental sense. In other words, there may exist (in addition to a minimal length), a {\it minimum norm}: states of sufficiently small norm do not exist in the Hilbert space. 

In \cite{BHZ1}, Equation (\ref{snap}) was described as a {\it snap-to} rule, as in snap to nearest site in discrete lattice (see Figure 1 for a qubit example). It could also be described as rounding the coefficient $c_n$ in (\ref{psisum}) to some finite accuracy in some chosen basis $\vert n \rangle$.

\begin{figure}[t!]
\label{bloch}
\includegraphics[width=4cm,height=4cm]{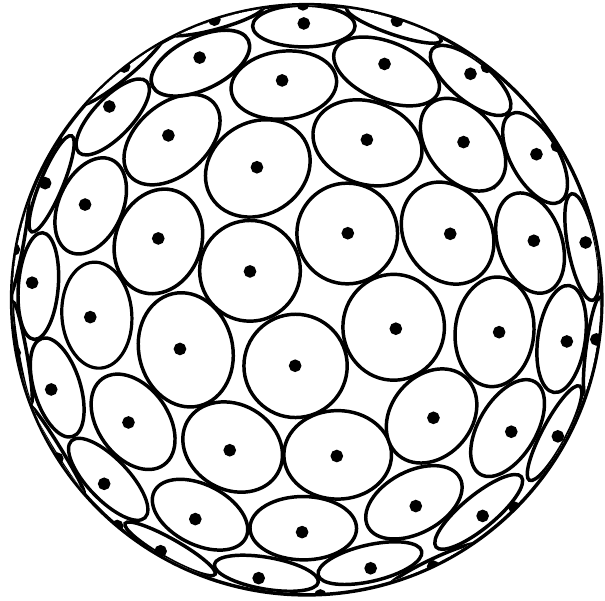}
\caption{A possible discretization of the Bloch sphere.}
\end{figure}

\section{Quantum Gravity and Decoherent Histories}

We can sketch, at least at the formal level, a version of quantum gravity that realizes the features described above. Let the partition function be a discrete Euclidean sum (see, e.g., \cite{Hartle,Ambjorn,Ambjorn:1998xu,Ambjorn:2011cg})
\begin{equation}
Z = \sum e^{- S( {\cal M}, \phi ) } 
\end{equation}
where $\cal M$ represents 4-geometries and $\phi$ matter field configurations. We assume an ultraviolet regulator -- for example, the geometries might be constructed from simplexes with a fundamental length $a$ which is non-zero but of order the Planck length (or possibly smaller). The configuration space, being discrete, is finite, and consequently the sum as well (e.g., assuming closed universes with finite topology). Of course there are many open questions related to this formulation, such as technical details of the triangulation, signature of the metric and causal structure, the phase diagram, long distance behavior, the emergence of symmetries such as general covariance, etc. Nevertheless, this class of quantum gravity models helps to illustrate properties of interest.

In this path integral approach, the amplitude connecting two configurations with (spacelike) 3-geometries ${\cal M}_3$ and matter state $\phi$ is
\begin{equation}
{\cal A} = \int D[ {\cal M} , \phi] ~e^{iS( {\cal M}, \phi )}
\end{equation}
where the 4-geometry ${\cal M}$ interpolates between the initial and final ${\cal M}_3$, and similarly for the matter fields. Again, the formal expression given above can be made concrete using a specific discretization of spacetime. Note, the initial and final states ${\cal M}_3$ in the amplitude ${\cal A}$ are eigenstates of geometry, so discretization of ${\cal M}$ (the 4-geometry to which the 3-geometries connect) implies a discrete set of basis states for 3-geometries. 

Further, a discrete simplexification or triangulation of the interpolating 4-geometry $\cal M$ implies that it has a minimum temporal extent. In other words, a minimum time step $a$ separates the initial and final configurations in the amplitude ${\cal A}$. Consider this smallest possible time evolution of a state: $\psi' = \exp(-iHa) \psi$. Then
\begin{equation}
\label{Ea}
\vert \psi' - \psi \vert \approx \vert \, -iHa \vert \psi \rangle \, \vert = a \langle \psi \vert H^2 \vert \psi \rangle^{1/2} \equiv {\cal E} \,a     ~~~,
\end{equation}
where ${\cal E} = 0$ only if $H \vert \psi \rangle = 0$. Otherwise, for a finite set of basis states $\psi$, there exists some smallest nonzero ${\cal E}_{\rm min}$ (presumably of order one over the size of the system $\sim L^{-1}$). Therefore, the smallest possible evolution causes either no change in $\psi$, or a change of distance at least ${\cal E}_{\rm min} a$. This defines a minimum norm, as states closer to $\psi$ than this distance are unnecessary to characterize the evolution of the system. Again, we see that a discrete version of the conventional continuous Hilbert space suffices to describe physics. A minimum norm, together with the snap-to rule in Equation (\ref{snap}), would be consistent with (\ref{Ea}).

Note, the continuum limit $a \rightarrow 0$ is sometimes considered in lattice studies of quantum gravity (or of gauge theory). A second order phase transition in this limit allows for infinite correlation length, making long distance physics independent of $a$ as it approaches zero. However, there is no physical requirement that $a$ be taken to exactly zero. In the case of lattice gauge theory we do not expect that the gauge fields live in a continuous spacetime -- short distance fluctuations of the metric intrude already at the Planck length. Similarly, the ultimate formulation of quantum gravity may contain a fundamental scale, presumably related to the Planck length, as indicated already in classical general relativity \cite{Calmet:2004mp}. (Such indications appear in Loop Quantum Gravity \cite{Rovelli}.) In any case, it seems that experiments inside the system (universe) will not be able to differentiate between models which impose the limit $a \rightarrow 0$, versus models with some finite but small $a$.

In quantum gravity, due to general covariance, the fundamental large scale objects are 4-geometries endowed with matter fields. (This does not seem to depend on whether the short distance description is a spacetime foam or fluctuating strings...) The considerations above suggest that maximally fine-grained decoherent histories may have a discrete character. While this has been suspected for some time with respect to the structure of spacetime itself, the implications for Hilbert space (i.e., the state space of quantum mechanics) have not been much explored. Because the minimum norm deduced for Hilbert space is so small, it hardly affects realistic laboratory experiments. However, it may have more fundamental implications for the Born rule and nature of the quantum multiverse, as we discuss below.

\section{Minimum Norm and Maverick Branches: Improbability Budget}

Hugh Everett \cite{Everett,Everett1,Hsu1,Hsu2} proposed that quantum measurement need not involve collapse (non-unitary projection) of the wavefunction. We briefly review this {\it no collapse} or {\it many worlds} formulation of quantum mechanics. Let $S$ be a single qubit and $M$ a device which measures the spin of the qubit along a particular axis. The corresponding eigenstates of spin are denoted $\vert \pm \rangle$. Define the operation of $M$ as follows
\begin{eqnarray}
\vert + \rangle \otimes \vert M \rangle ~\longrightarrow ~  \vert + \, , \, M_+ \rangle  \nonumber \\
\vert - \rangle \otimes \vert M \rangle ~\longrightarrow ~  \vert - \,,\, M_- \rangle 
\end{eqnarray}
where $M_\pm$ is the state of the apparatus after recording a $\pm$ outcome. What happens to a superposition state $\vert \Psi_S \rangle =  c_+ \vert + \rangle ~+~ c_- \vert - \rangle$? In the conventional formulation, with measurement collapse, {\it one} of the two final states $\vert + \, , \, M_+ \rangle~$ {\it or} $~\vert - \, , \, M_- \rangle$ results, with probabilities $\vert c_+ \vert^2$ and $\vert c_- \vert^2$ respectively. The probability rule is the Born Rule, which enters the conventional formulation together with wavefunction collapse.

However, if the combined system $S' = S + M$ evolves unitarily (in particular, {\it linearly}): $\Psi_{S'} = \exp( - i H t ) \Psi_{S'}$, we obtain a superposition of measurement device states:
\begin{equation}
\big(  c_+ \vert + \rangle ~+~ c_- \vert - \rangle  \big ) \otimes \vert M \rangle ~~ \longrightarrow ~~
c_+ \, \vert + \, , \, M_+ \rangle ~+~ c_- \, \vert - \, , \, M_- \rangle ~~.
\end{equation}
This seems counter to our experience: measurements produce a single outcome, not a superposition state. However, as noted by Everett, an observer in state $M_+$ is unaware of the other state $M_-$ due to decoherence \cite{decoherence,decoherence1}. First, a semi-classical measuring device or observer will have many degrees of freedom. Second, a measurement requires that the states $M_+$ and $M_-$ differ radically: the outcome must be stored redundantly, and accessible to macroscopic beings. Therefore, the overlap of states $M_+$ and $M_-$ must be $\exp( - N )$, where $N$ is a large number of degrees of freedom, and (one assumes; see below) the future dynamical evolution of each branch is unlikely to alter this situation. For All Practical Purposes (J.S. Bell \cite{Bell}), the branches have decohered from each other. Each observer perceives a collapse and a distinct outcome, however the evolution of $S'$ is deterministic and linear, governed by the Schrodinger equation.

Decoherence shows that quantum measurement can proceed continuously, as different branches rapidly lose contact with each other. The result is the {\it appearance} of collapse (a single outcome) to each observer $M_{\pm}$ within $S'$. A no collapse universe is intrinsically quantum, containing all possible branches, but in which observers with discordant memory records are unaware of each other.

Consider $N$ qubits: $\Psi = \otimes_{i=1}^N ~ \psi_i$, with each prepared in the identical state $\psi_i = c_+ \vert + \rangle_i ~+~ c_- \vert - \rangle_i$, with $c_\pm$ non-zero. Unitary evolution implies that all possibilities are realized, including, e.g., all spins are measured in the $+$ state: $\Psi \sim \vert + + + \cdots + \rangle$. If $\vert c_+ \vert$ is small, then this outcome is very unlikely according to the Born Rule. But for $c_+$ not exactly zero, it remains one of $2^N$ distinct possible outcomes. Each outcome implies the existence of an observer with their own distinct memory records. 

For $N$ sufficiently large and $|c_+| \neq |c_-|$, it can be shown \cite{BHZ2} that the vast majority of the $2^N$ realized observers (i.e., counting each distinct observer equally) see an outcome which is highly unlikely according to the Born probability rule. Counting possible outcomes is simply combinatorial and the result does not depend on the values $c_\pm$. As $N \rightarrow \infty$, for all non-zero values of $c_\pm$, almost all of the realized observers find nearly equal number of $+$ and $-$ spins. In contrast, the Born rule predicts that the relative number of $+$ and $-$ outcomes depends on $|c_\pm|^2$.

Clearly it is a challenge to explain why a specific observer (i.e., you or me) in the quantum multiverse finds evidence for the Born Rule. This problem was well known to Everett. He called small norm branches of the wavefunction {\it maverick branches}. The difficulty is to explain why we do not ourselves reside on a maverick branch, without imposing (in a circular fashion) the Born Rule a priori. It is important to note that proper functioning of the measuring device $M$, the approach to thermal (statistical) equilibrium, and even the emergence of a semi-classical reality (spacetime metric, classical fields, persistent objects) depend on the Born Rule. We will return to these issues below. For convenience, we will sometimes abbreviate the Born Rule as BR. Terms like probable or improbable will be understood to mean with respect to the BR or norm squared measure in Hilbert space, with the understanding that the no collapse version of quantum mechanics does not, by itself, impose any specific probability measure.

The existence of a {\it minimum norm} removes the most improbable maverick states. For a given minimum norm we can characterize the degree of deviation from outcomes which are most likely under the usual BR. Would these deviations be detectable to observers on the most improbable branches that remain (i.e., those just above the threshold of minimum norm)? If the answer is NO (for macroscopic observers like ourselves), then we will have answered the question of why physicists in a many worlds multiverse believe in (observe the) Born Rule, even though it cannot be imposed directly in a no collapse version of quantum mechanics.

In \cite{BHZ2} the $N$ qubit state $\Psi$ given above was studied under minimum norm assumption: $\vert \psi - \psi' \vert < \epsilon$ and $\vert \Psi - \Psi' \vert < \sqrt{N} \epsilon$, with $N \epsilon^2$ small. For given $N$ and $\epsilon$ the expected deviation in frequency of $+$ outcomes is of order $\vert \ln N \epsilon^2 \vert^2$ standard deviations. For branches that are just at the threshold of minimum norm, we can think in terms of an {\it improbability budget}: the distribution of improbable outcomes that cause the norm of the state to be small. Are these improbable outcomes detectable?

Importantly, a specific observer can only measure a small fraction of the total number of decoherent outcomes which define their branch of the wavefunction. Suppose the observer can measure $n$ outcomes from $\Psi$, with statistical uncertainty $n^{-1/2}$ much larger than $N^{-1/2}$. If 
\begin{equation}
\label{SD}
n^{-1/2} ~\gg~ \vert \ln N \epsilon^2 \vert^2 \, N^{-1/2}
\end{equation}
the observer will fail to detect the anomalous improbability. Indeed, the process of decoherence itself, and of approach to statistical equilibrium, both involve probabilistic behavior. Vast improbabilities can been hidden in, e.g., the evolution of systems towards decoherence or equilibrium, which take place at all times, but are generally not observed or closely monitored. Thus we expect $n$ to be vanishingly small compared to $N$, and for the inequality above to hold.

Consider statistical equilibrium, which is a generic consequence of pure state evolution of closed systems: ``typical'' quantum pure states are highly entangled, and the density matrix describing any small sub-system (obtained by tracing over the rest of the pure state) is very close to micro-canonical (i.e., thermal) \cite{VN1,VN2}. Under dynamical (Schrodinger) evolution, all systems (even those that are initially far from typical) spend nearly all of their time in a typical state (modulo some weak conditions on the Hamiltonian). However, it is possible for a system to be in a typical pure state and yet (improbably, under BR) for its properties on some decoherent branches to deviate from thermal equilibrium values. A single gram of interstellar hydrogen, or nitrogen in Earth's atmosphere, could contribute enormously to the improbability budget through such deviations, without ever being detected by a physicist. 

Next, consider decoherence and the emergence of semi-classical reality. This is also a dynamical process, which incorporates statistical amplification of quantum outcomes, creating redundant correlations between the outcome state and a macroscopic number of environmental degrees of freedom \cite{NR}. One can imagine many examples of slight deviations from most probable (under assumption of BR) decoherence behavior that are nevertheless hard to detect. For example, a macroscopic system might decohere slower (or faster) than predicted under BR. A deviation in either direction of a few nanoseconds (e.g., for a gram of matter) might be enormously improbable, but very difficult to detect. Such decoherence processes are happening at all times, but almost none are under laboratory observation.

These examples illustrate that nearly all of the $N$ potential decoherent outcomes defining our branch of the multiverse are beyond close monitoring, leading to an infinitesimal ratio of $n$ versus $N$. Hence, Equation (\ref{SD}) is satisfied and macroscopic observers will not detect deviation from the Born Rule after removal of the smallest norm maverick branches.

\section{Conclusions}

No collapse (or many worlds) versions of quantum mechanics are often characterized as extravagant, because of the many branches of the wavefunction. However it is also extravagant to postulate that spacetime or Hilbert space are infinitely continuous. Continuous Hilbert space requires that for any two choices of orientation of a qubit spin (see Figure 1), no matter how close together, there are an infinite number of physically distinct states between them, with intermediate orientation. Instead, there may only be a finite (but very large) number of distinct orientations allowed, suggesting a minimum norm in Hilbert space. No experiment can probe absolute continuity, and indeed there seem to be fundamental limits on such experiments, arising from quantum gravity itself.

We illustrated a direct connection between discrete spacetime (the simplex length $a$) and discrete Hilbert space (minimum non-zero distance in Hilbert space produced by time evolution), in a specific class of quantum gravity models based on Feynman path integrals. It may be the case that maximally fine-grained decoherent histories generated within quantum gravity have discrete geometries and exist in a discrete Hilbert space. Consequently histories with sufficiently small norm are never generated, thereby solving Everett's problem with maverick branches. In the remaining branches, deviations from Born Rule probabilities are almost entirely hidden from semi-classical observers. 

A final comment on unitarity. Conventionally, one assumes (in the absence of measurement collapse) a continuous Hilbert space and absolute unitarity in the evolution of the wavefunction. In a discrete Hilbert space, where some branches of the (continuous) wavefunction with very small norm (Born Rule probability) are excluded, unitarity must be slightly violated. However, viewed from the spacetime (4-geometry) perspective, this is simply a restriction on the set of decoherent histories that are realized, presumably a consequence of quantum gravity dynamics with a discrete character. Note the restriction is very weak: the histories eliminated are those with the smallest norm, and the minimum norm appears naturally as a consequence of a deep property of quantum gravity -- fundamental discreteness at the Planck scale.

An extreme version of the idea that only a subset of possible decoherent histories is realized, is that only {\it one} decoherent history is ``real''  \cite{Csonka,GMH}. This proposal, explored by Gell-Mann and Hartle after decades of work on decoherent histories \cite{GMH}, is meant to avoid the perceived extravagance of many worlds. Of course, the assumption implies maximal violation of unitarity: only one outcome is realized per decoherence event, as in Copenhagen with its von Neumann projection postulate. From the spacetime perspective this universe (the single decoherent history) is entirely deterministic, with only the {\it appearance} of quantum randomness to beings inside.

\bigskip

\newpage

\end{document}